\documentclass[aps,superscriptaddress,twocolumn,floatfix,pra,nofootinbib,a4paper]{revtex4-2}
\usepackage{times}
\usepackage{amsfonts}
\usepackage{amsmath}
\usepackage{amssymb}
\usepackage{amsthm}
\usepackage{color}
\usepackage{multirow}
\usepackage[normalem]{ulem}
\usepackage{natbib}
\usepackage[T1]{fontenc}
\usepackage{float}
\usepackage{mathtools}
\usepackage{bm}
\usepackage{graphicx}
\usepackage{xcolor}
	\definecolor{carmine}{RGB}{150,0,24}
\usepackage[colorlinks=true,linkcolor=blue,citecolor=magenta,urlcolor=blue]{hyperref}

\allowdisplaybreaks[4]
\DeclareMathOperator{\Tr}{tr}

\newcommand{\ket}[1]{|#1\rangle}

\newcommand{\bracket}[3]{\langle#1|#2|#3\rangle}
\newcommand{\ketbra}[2]{|#1\rangle\langle#2|}
\newcommand{\expect}[1]{\langle#1\rangle}

\newcommand{\norm}[1]{\left\lVert#1\right\rVert}

\begin{document}


\title{Entanglement detection with imprecise measurements}

\author{Simon Morelli}\thanks{S.M. and H.Y. contributed equally to this manuscript.}
\author{Hayata Yamasaki}\thanks{S.M. and H.Y. contributed equally to this manuscript.}
\author{Marcus Huber}
\author{Armin Tavakoli}
\affiliation{Institute for Quantum Optics and Quantum Information -- IQOQI Vienna, Austrian Academy of Sciences, Boltzmanngasse 3, 1090 Vienna, Austria}
\affiliation{Atominstitut,  Technische  Universit{\"a}t  Wien, Stadionallee 2, 1020  Vienna,  Austria}

\begin{abstract}
	We investigate entanglement detection when the local measurements only nearly correspond to those intended. This corresponds to a scenario in which measurement devices are not perfectly controlled, but nevertheless operate with bounded inaccuracy. We formalise this through an operational notion of inaccuracy that can be estimated directly in the lab. To demonstrate the relevance of this approach,  we show that small magnitudes of inaccuracy can significantly compromise several well-known entanglement witnesses. For two arbitrary-dimensional systems, we show how to compute tight corrections to a family of standard entanglement witnesses due to any given level of measurement inaccuracy. We also develop semidefinite programming methods to bound correlations in these scenarios.
\end{abstract}

\maketitle

\textit{Introduction.---}  Deciding whether an initially unknown state is entangled is one of the central challenges of quantum information science \cite{Guhne2009, Horodecki2009, Friis2019}. The most common approach is the method of entanglement witnesses, in which one hypothesises that the state is close to a known target and then finds suitable local measurements that can reveal its entanglement \cite{Horodecki1996, Terhal1999, Lewenstein2000}. In principle, this allows for the detection of every entangled state. However, it crucially requires the experimenter to  flawlessly perform the stipulated quantum measurements. This  is an idealisation to which one may only aspire: even for the simplest system of two qubits, small alignment errors can cause false positives \cite{Seevinck2007, Rosset2012}. In contrast, by adopting a device-independent approach, any concerns about the modelling of the measurement devices can be dispelled. This entails viewing them as quantum black boxes  and detecting entanglement through the violation of a Bell inequality \cite{Bancal2011, Moroder2013}. However, Bell experiments are practically demanding \cite{Brunner2014}. Also, many entangled states either cannot, or are not known to, violate any Bell inequality \cite{Werner1989, Augusiak2014}. In addition, for the common purpose of verifying that a non-malicious entanglement source operates as intended, a device-independent approach is to use a sledgehammer to crack a nut. In the interest of a compromise, entanglement detection has also been investigated in steering scenarios, in which some devices are assumed to be perfectly controlled and others are quantum black boxes \cite{Wiseman2007}. Nevertheless, such asymmetry is often not present in non-malicious scenarios, and the approach still suffers from drawbacks similar to both the device-independent case, albeit it milder, and the standard, fully controlled, scenario. A much less explored compromise route is to only assume knowledge of the Hilbert space dimension \cite{Moroder2012, Tavakoli2018}. This essentially adopts the view that the experimenter has no control over the relevant degrees of freedom. Such ideas have also been used to strengthen steering-based entanglement detection \cite{Moroder2016}.

Here, we introduce an approach to entanglement detection that neither assumes flawless control of the measurements nor views them as mostly uncontrolled operations. The main idea is that an experimenter can quantitatively estimate the accuracy of their measurement devices and then base entanglement detection on this  benchmark. Such knowledge naturally requires a fixed Hilbert space dimension: the experimenter knows the degrees of freedom on which they operate. 
To quantify the inaccuracy between the intended target measurement and the lab measurement, we use a simple fidelity-based notion that can handily be measured experimentally. 

In what follows, we first establish the relevance of small inaccuracies by showcasing that the conclusions of well-known entanglement witnesses can be substantially compromised. We show that the magnitude of detrimental influence associated to a small inaccuracy does not have to decrease for higher-dimensional systems. This is important because higher-dimensional entangled systems are increasingly interesting for experiments  \cite{Dada2011, Erhard2020, Ecker2019, Herrera2020, Hu2020} but typically cannot be controlled as precisely  as qubits. Secondly, we develop entanglement criteria that explicitly take the degree of inaccuracy into account. For two-qubit scenarios, we provide this based on the simplest entanglement witness and the Clauser-Horne-Shimony-Holt (CHSH) quantity.  For a pair of systems of any given local dimension, we show that such criteria can be analytically established as corrections to a simple family of standard entanglement witnesses. 
Finally, we present semidefinite programming (SDP) relaxations for bounding the set of quantum correlations under measurement inaccuracies. We use this both to estimate the potentially constructive influence of measurement inaccuracy on entanglement-based correlations and to systematically place upper bounds for separable states on linear witnesses.


\textit{Framework.---} We consider sources of bipartite states $\rho=\rho_\text{AB}$ of local dimension $d$. The subsystems are measured individually with settings $x$ and $y$ respectively, producing outcomes $a,b\in\{1,\ldots,o\}$. The experimenter's aim is to measure the first (second) system  using a set of projective measurements $\{\tilde{A}_{a|x}\}$ ($\{\tilde{B}_{b|y}\}$). These are called target measurements. However, the measurements actually performed in the lab do not precisely correspond to the targeted measurements, but instead to positive operator-valued measures (POVMs) $\{A_{a|x}\}$ ($\{B_{b|y}\}$). These are called lab measurements and do not need to be projective.  The correlations in the experiment are given by the Born-rule
\begin{equation}\label{born}
p(a,b|x,y)=\Tr\left[A_{a|x}\otimes B_{b|y}\rho\right].
\end{equation}
We quantify the correspondence between each of the target measurements and the associated lab measurements through their average fidelity, 
\begin{align}\label{fidelity}
& \mathcal{F}^\text{A}_x\equiv \frac{1}{d}\sum_{a=1}^{o}\Tr\left[A_{a|x} \tilde{A}_{a|x} \right],
& \mathcal{F}^{\text{B}}_y\equiv \frac{1}{d}\sum_{b=1}^{o}\Tr\left[B_{b|y} \tilde{B}_{b|y} \right].
\end{align}
The fidelity respects $\mathcal{F}\in[0,1]$ with $\mathcal{F}=1$ if and only if the lab measurement is identical to the target measurement. Importantly, the fidelity admits a simple operational interpretation: it is the average probability of obtaining outcome $a$ ($b$) when the lab measurement is applied to each of the orthonormal states spanning the eigenspace of the $a$-th ($b$-th) target projector. Thus, the fidelities $\{\mathcal{F}^\text{A}_x,\mathcal{F}^\text{B}_y\}$ can be directly determined by probing the lab measurements with single qudits from a well-calibrated, auxiliary, source. This requires no entanglement and can routinely be achieved, see e.g.~Ref.~\cite{Bouchard2018}. It motivates the assumption of a bounded inaccuracy, i.e.~a lower bound on each of the fidelities,
\begin{align}\label{assumption}
& \mathcal{F}^\text{A}_x\geq 1-\varepsilon^\text{A}_x, &  \mathcal{F}^\text{B}_y\geq 1-\varepsilon^\text{B}_y,
\end{align}
where the parameter $\varepsilon\in[0,1]$ is the inaccuracy of the considered lab measurement. In the extreme case of $\varepsilon=0$, the lab measurement is identical to the target measurement and our scenario reduces to a standard entanglement witness. In the other extreme, $\varepsilon=1$, only the Hilbert space dimension of the measurement is known. Away from these extremes, one encounters the more realistic scenario, in which the experimenter knows the degrees of freedom, but is only able to control them up to a limited accuracy. 

The simplest tests of entanglement use the minimal number of outcomes ($o=2$). In such scenarios the fidelity constrains \eqref{assumption} can be simplified into  
\begin{align}\label{observable}
& \Tr\left(A_{x}\tilde{A}_{x}\right)\geq d\left(1-2\varepsilon^\text{A}_x\right), &&\Tr\left(B_{y}\tilde{B}_{y}\right)\geq d\left(1-2\varepsilon^\text{B}_y\right)
\end{align}
where we have defined observables $A_{x}\equiv A_{1|x}-A_{2|x}$ and $B_{y}\equiv B_{1|y}-B_{2|y}$. The observables can be arbitrary Hermitian operators whose extremal eigenvalue is bounded by unity, i.e.~$\norm{A_x}_\infty\leq 1$ and $\norm{B_y}_\infty\leq 1$. 

Notice that the proposed framework immediately extends also to multipartite scenarios.

\textit{Impact of inaccuracies in entanglement witnessing.---} A crucial motivating question for our approach is whether, and to what extent, small inaccuracies  in the measurement devices ($\varepsilon\ll 1$) impact the analysis of a conventional entanglement witness. We discuss this matter based on several well-known witnesses.

Firstly, consider the simplest entanglement witness for two qubits, involving two pairs local Pauli observables: $\mathcal{W}=\expect{\sigma_X\otimes\sigma_X}+\expect{\sigma_Z\otimes\sigma_Z}$. For separable states we have $\mathcal{W}\leq \mathcal{W}_\text{sep}=1$ and for entangled states $\mathcal{W}\leq \mathcal{W}_\text{ent}=2$. Consider now that the lab observables $\{A_1,A_2\}$ and $\{B_1,B_2\}$ only nearly correspond \eqref{observable} to the target observables $\{\sigma_X,\sigma_Z\}$. Since $\mathcal{W}_\text{ent}=2$ is algebraically maximal, it remains unchanged, but such is not the case for the separable bound $\mathcal{W}_\text{sep}$. Thanks to the simplicity of $\mathcal{W}$, we can precisely evaluate $\mathcal{W}_\text{sep}$ in the prevalent scenario when all measurement devices are equally inaccurate, i.e.~$\varepsilon^\text{A}_x=\varepsilon^\text{B}_y=\varepsilon$. For a product state, we have $\mathcal{W}=\expect{A_1}\expect{B_1}+\expect{A_2}\expect{B_2}\leq 
\sqrt{\expect{A_1}^2+\expect{A_2}^2} \sqrt{\expect{B_1}^2+\expect{B_2}^2}$. Since the target measurements are identical on both sites and the factors are independent, they are optimally chosen equal. Then, it is easily shown  that the optimal choice of Bloch vectors corresponds to aligning $A_1$ and $A_2$ ($B_1$ and $B_2$) to the extent allowed by $\varepsilon$. This leads to the following tight condition for entanglement detection (see Supplementary Material)
\begin{equation}\label{qubit}
\mathcal{W}_\text{sep}(\varepsilon)= 1+4\left(1-2\varepsilon\right)\sqrt{\varepsilon\left(1-\varepsilon\right)},
\end{equation}
when $\varepsilon\leq \frac{1}{2}-\frac{1}{2\sqrt{2}}$ and $\mathcal{W}_\text{sep}=2$ otherwise. Importantly, the derivative diverges at $\varepsilon\rightarrow 0^+$. Hence, a small $\varepsilon$ induces a large perturbation in the ideal ($\varepsilon=0$) separable bound.  In the vicinity of $\varepsilon=0$, it scales as $\mathcal{W}_\text{sep}\sim 1+4\sqrt{\varepsilon}$. For example, $\varepsilon=0.5\%$ leads to $\mathcal{W}_\text{sep}(\varepsilon)\approx 1.28$, which eliminates over a quarter of the range in which  standard entanglement detection is possible, indicating the relevance of false positives.

Secondly, consider  the CHSH quantity for entanglement detection, namely $\mathcal{W}=\expect{\sigma_X\otimes \left(\sigma_X+\sigma_Z\right)}+\expect{\sigma_Z\otimes \left(\sigma_X-\sigma_Z\right)}$. Here, we have targeted observables optimal for violating the CHSH Bell inequality \cite{CHSH1969}. One has $\mathcal{W}_\text{sep}=\sqrt{2}$ and $\mathcal{W}_\text{ent}=2\sqrt{2}$. In contrast to the previous example, the fact that all correlations from $d$-dimensional separable states constitute a subset of all correlations based on local hidden variables implies that entanglement can be detected for any value of $\varepsilon$. However, as we show in Supplementary Material through an explicit separable model that we conjecture to be optimal, this fact does not qualitatively improve the robustness of idealised ($\varepsilon=0$) entanglement detection to small inaccuracies. We obtain
\begin{align}
\mathcal{W}_\text{sep}=4\left(1-2\varepsilon\right)\sqrt{\varepsilon(1-\varepsilon)}+\sqrt{2-16\varepsilon\left(1-\varepsilon\right)\left(1-2\varepsilon\right)^2},
\end{align}
when $\varepsilon\leq \frac{1}{2}-\frac{1}{2\sqrt{2}}$ and $\mathcal{W}_\text{sep}=2$ otherwise. For small $\varepsilon$, we find $\mathcal{W}_\text{sep}\sim \sqrt{2}+4\sqrt{\varepsilon}$. An inaccuracy of $\varepsilon=0.5\%$ ensures $\mathcal{W}_\text{sep}\gtrsim 1.67 $, which eliminates nearly a fifth of the range in which  standard entanglement detection is possible.

\begin{figure}[t!]
	\centering
	\includegraphics[width=\columnwidth]{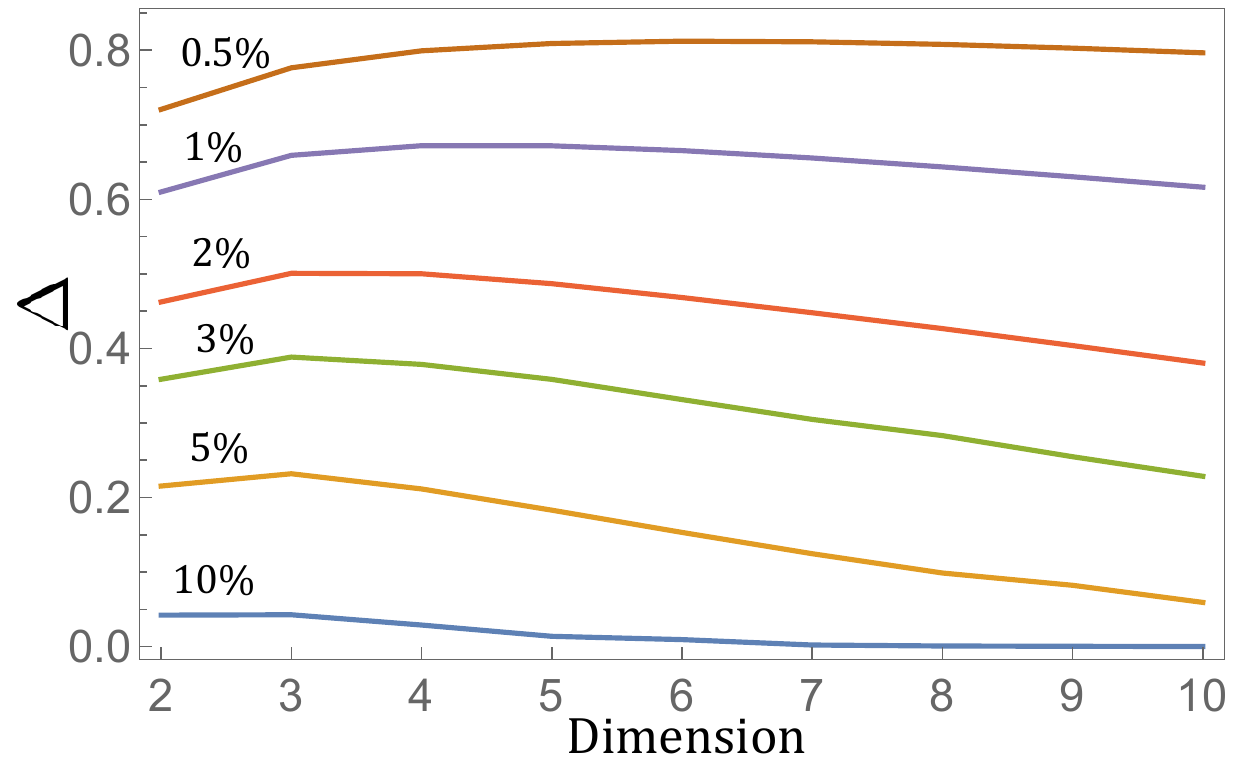}
	\caption{Numerically obtained lower bounds on the relative magnitude of the entangled-to-separable gap, $\Delta$, for entanglement witnessing based on two conjugate bases at  different degrees of measurement inaccuracy $\varepsilon\in\{0.5\%, 1\%, 2\%, 3\%, 5\%, 10\%\}$. }\label{FigNumerics}
\end{figure}

Interestingly, it is \textit{a priori} not clear how small $\varepsilon$ should impact standard entanglement witnessing as $d$ increases.  On the one hand, the impact ought to increase due to the increasing number of orthogonal directions in Hilbert space. On the other hand, it ought to decrease due to the growing distances in Hilbert space. For instance, the $\varepsilon$ required to transform the computational basis into its Fourier transform scales as $\varepsilon=\frac{\sqrt{d}-1}{\sqrt{d}}$, which rapidly approaches unity. To investigate the trade-off between these two effects, we consider the $d$-dimensional generalisation of the simplest entanglement witness. Both subsystems are subject to the same pair of target measurements, namely the computational basis $\{\ket{e_i}\}_{i=1}^d$ and its Fourier transform $\{\ket{f_i}\}_{i=1}^d$, where $\ket{f_i}=\Omega\ket{e_i}$ with $\Omega_{jk}=\frac{1}{\sqrt{d}}e^{\frac{2\pi i}{d}jk}$. The witness is 
$\mathcal{W}^{(d)}=\sum_{i=1}^d \bracket{e_i,e_i}{\rho}{e_i,e_i}+\bracket{f_i,f_i}{\rho}{f_i,f_i}$. Notice that for $d=2$ this only differs from the previous, simplest, witness by a normalisation term. One has  $\mathcal{W}^{(d)}_\text{sep}=1+\frac{1}{d}$ and $\mathcal{W}^{(d)}_\text{ent}=2$ \cite{Spengler2012}.  Allowing for measurement inaccuracy, we use an alternating convex search algorithm to numerically optimise over the lab measurements and shared separable states to obtain lower bounds on $\mathcal{W}^{(d)}_\text{sep}(\varepsilon)$. See Supplementary Material for details about the method. In order to compare the impact of measurement inaccuracy for different dimensions, we consider the  following ratio between the entangled-to-separable gap in the inaccurate and ideal case, $\Delta\equiv \frac{\mathcal{W}^{(d)}_\text{ent}(0)-\mathcal{W}^{(d)}_\text{sep}(\varepsilon)}{\mathcal{W}^{(d)}_\text{ent}(0)-\mathcal{W}^{(d)}_\text{sep}(0)}=\frac{d}{d-1}\left[2-\mathcal{W}^{(d)}_\text{sep}(\varepsilon)\right]$. Notice that the numerator features $\mathcal{W}^{(d)}_\text{ent}(0)$ instead of $\mathcal{W}^{(d)}_\text{ent}(\varepsilon)$ because $\varepsilon$ is not in itself a resource for the experimenter. The results of the numerics are illustrated in Figure~\ref{FigNumerics} for some different choices of $\varepsilon$. We observe that $\Delta$ is not  monotonic in $d$, but instead features a maximum, that shifts downwards in $d$ as $\varepsilon$ increases. Beyond this maximum point, the impact of measurement inaccuracies grows as the dimension becomes large.

Finally, for multipartite qubit states, it is natural to expect that  the detrimental influence of small $\varepsilon$ grows with the number of qubits under consideration. The reason is that measurement inaccuracies can accumulate separately in the different subsystems. This intuition is confirmed by the models of Ref.~\cite{Rosset2012}, in which small alignment errors are used to spoof, with increasing magnitude, the standard fidelity-based witness of genuine multipartite entanglement for  Greenberger-Horne-Zeilinger states \cite{Bourennane2004}. 
This further confirms the need of considering measurement inaccuracies.

\textit{High-dimensional entanglement criterion.---} In view of the the relevance of small measurement inaccuracies, it is natural to formulate entanglement criteria that take them explicitly into account beyond the simplest, two-qubit, scenario.  Consider a pair of $d$-dimensional systems and $n\in\{1,\ldots,d^2-1\}$ measurements. For system A, the observables ideally correspond to (subsets of) a  generalised Bloch basis $\{\lambda_i\}_{i=1}^{n}$ and for system B, the ideal observables are the complex conjugates $\{\bar{\lambda}_i\}_{i=1}^{n}$. Here, $\lambda_i$ is $d$-dimensional, traceless and satisfies $\Tr\left(\lambda_i\lambda_j^\dagger\right)=d\delta_{ij}$ \cite{Bertlmann2008}. Defining $\rho=\frac{1}{d}\left(\openone+\sum_{i=1}^{d^2-1}\mu_i\lambda_i\right)$, one has $\norm{\vec{\mu}}^2\leq d-1$. A simple standard entanglement witness, based on a total of $n$ measurements, is then given by 
\begin{equation}\label{oldwitness}
\mathcal{W}^{(d)}=\sum_{i=1}^n \expect{\lambda_i\otimes \bar{\lambda}_i}.
\end{equation}
Using H\"older's inequality, one finds that separable states obey  $\mathcal{W}^{(d)}_\text{sep}=d-1$. When the choice of Bloch basis is fixed, entangled states can achieve at most $\mathcal{W}_\text{ent}^{(d)}=\nu_\text{max}\left[\sum_{i=1}^n \lambda_i\otimes \bar{\lambda}_i\right]$, by choosing the state as the eigenvector corresponding to the largest eigenvalue ($\nu_\text{max}$). When the choice of Bloch basis is not fixed, a general upper bound for entanged states is $\mathcal{W}^{(d)}_\text{ent}\leq \min\left\{\sqrt{n\left(d^2-1\right)},n(d-1)\right\}$, as shown in Supplementary Material. Note that $n(d-1)$ only is relevant when $d=2$. Notice also that the maximally entangled state $\ket{\phi^+_d}=\frac{1}{\sqrt{d}}\sum_{i=0}^{d-1}\ket{ii}$ achieves $\mathcal{W}^{(d)}=n$ regardless of the choice of Bloch basis. 

Consider now that the lab observables only nearly correspond to $\{\lambda_i\}$ and $\{\bar{\lambda}_i\}$ respectively. We write them as
$A_i=q \lambda_i+\sqrt{1-q^2}\lambda_{i}^{\perp}$ and $B_i=q \bar{\lambda}_i+\sqrt{1-q^2}\bar{\lambda}_{i}^{\perp}$, where $q\in[-1,1]$ is related to the inaccuracy through $q=1-2\varepsilon$ and $\lambda_{i}^{\perp}$ and $\bar{\lambda}_{i}^{\perp}$ are observables orthogonal to $\lambda_i$ and $\bar{\lambda}_i$, respectively, on the generalised Bloch sphere. In Supplementary Material, we prove that the witness $\mathcal{W}^{(d)}=\sum_{i=1}^n\expect{A_i\otimes B_i}$ for separable states obeys 
\begin{equation}\label{bound}
\mathcal{W}^{(d)}_\text{sep}(\varepsilon)\leq \left(d-1\right) \left(q+\sqrt{n-1}\sqrt{1-q^2}\right)^2,
\end{equation}
when $q\geq \frac{1}{\sqrt{n}}$ and otherwise $\mathcal{W}^{(d)}_\text{sep}(\varepsilon)\leq n\left(d-1\right)$, which is algebraically maximal. As is intuitive, the window for detecting entanglement shrinks as $\varepsilon$ increases.

We investigate the tightness of the bound. To this end, choose the state as $\ket{\phi^\dagger}\otimes \ket{\phi^T}$, where the local Bloch vector is $\mu_i=\frac{\sqrt{d-1}}{\sqrt{n}}$ and where $\lambda_i\rightarrow \lambda_i^\dagger$ ($\lambda_i\rightarrow \lambda_i^T$) for $\ket{\phi^\dagger}$ ($\ket{\phi^T}$). Choose the observables as $A_i=q\lambda_i+\sum_{j\neq i}\frac{\sqrt{1-q^2}}{\sqrt{n-1}}\lambda_j$ and $B_i=q\bar{\lambda}_i+\sum_{j\neq i}\frac{\sqrt{1-q^2}}{\sqrt{n-1}}\bar{\lambda}_j$. This returns the separable bound \eqref{bound}. However, we need to check that the Bloch vector $\vec{\mu}$ corresponds to a valid state. Curiously, for the most powerful case, namely $n=d^2-1$, tightness would be implied  by a positive answer to the long-standing open question of whether there exists a Weyl-Heisenberg covariant symmetric informationally complete (SIC) POVM in dimension $d$. To see the connection, simply choose the Bloch basis as the non-Hermitian Weyl-Heisenberg basis $\{X^uZ^v\}$ for $u,v\in\{0,\ldots,d-1\}$ and $u+v>0$, where  $X=\sum_{k=0}^{d-1}\ketbra{k+1}{k}$ and $Z=\sum_{k=0}^{d-1}e^{\frac{2\pi ik}{d}}\ketbra{k}{k}$. It follows immediately that $|\bracket{\phi}{X^uZ^v}{\phi}|=\frac{1}{\sqrt{d+1}}$, which defines a SIC-POVM. Since these SIC-POVMs are conjectured to exist in all dimensions \cite{Zauner2011}, and are known to exist up to well above the first hundred dimensions \cite{Scott2017, Fuchs2017}, our bound is plausibly tight for any $d$.

\textit{SDP methods.---} We develop a hierarchy of SDP relaxations to bound the largest possible value of any linear witness, $\mathcal{W}=\sum_{a,b,x,y} c_{abxy}p(a,b|x,y)$, for some real coefficients $c_{abxy}$. The method applies both for correlations originating from entangled states and from separable states, under any given degree of measurement inaccuracy and arbitrary target measurements. Thus, we systematically establish upper bounds $\mathcal{W}^\uparrow_\text{ent}(\varepsilon)\geq \mathcal{W}_\text{ent}(\varepsilon)$ and $\mathcal{W}^\uparrow_\text{sep}(\varepsilon)\geq \mathcal{W}_\text{sep}(\varepsilon)$. This has a three-fold motivation. Firstly, $\mathcal{W}_\text{ent}$ will generally depend on $\varepsilon$; cases with $\mathcal{W}^{(d)}>\mathcal{W}^{(d)}_\text{ent}(0)$ can be observed when the inaccuracies accumulate in a constructive way (e.g.~a favourable systematic error in the local reference frames). It is relevant to bound such occurances. Secondly, knowledge of $\mathcal{W}^\uparrow_\text{ent}(\varepsilon)$ allows an experimenter to give lower bounds on the inaccuracy of the measurement devices. Thirdly, and most importantly, this enables a general and systematic construction of entanglement witnesses of the form   $\mathcal{W}\leq \mathcal{W}^\uparrow_\text{sep}(\varepsilon)$.

We discuss the main features of the method for computing  $\mathcal{W}_\text{ent}^\uparrow(\varepsilon)$ and then see how it can be extended to also compute  $\mathcal{W}_\text{sep}^\uparrow(\varepsilon)$. To this end, as is standard, the SDP relaxation method is based on the positivity of a moment matrix. This matrix consists of traces of monomials (in the spirit of e.g.~\cite{Burgdorf2012}) which are composed of products of the state, the lab measurements and the target measurements (see Supplementary Material for specifics). Moments corresponding to products of the first two can be used to build a generic linear witness $\mathcal{W}$ via Eq.~\eqref{born}. Moments corresponding to products of the final two can be used to build the constraints on the fidelities $\mathcal{F}^\text{A}_x$ and $\mathcal{F}^\text{B}_y$. Our construction draws inspiration from two established ideas. Firstly, one can capture the constraints of $d$-dimensional Hilbert space, on the level of the moment matrix, by numerically sampling states and measurements \cite{Navascues2015}. Secondly, in scenarios without entanglement, constraints capturing the fidelity of a quantum state with a target can be incorporated into the moment matrix \cite{Tavakoli2021}. We adapt the latter to entanglement-based scenarios and measurement fidelities as needed for Eq.~\eqref{assumption}. Details are given in Supplementary Material. We have applied this method, at low relaxation level, in several different case studies in low dimensions and frequently found that the obtained upper bounds coincide with those obtained from interior point optimisation routines. We note that the computational requirements for this tool can be much reduced since sampling-based symmetrisation methods of Ref.~\cite{Tavakoli2019} can straightforwardly be incorporated.

To extend this method for the computation of $\mathcal{W}_\text{sep}^\uparrow(\varepsilon)$, we must incorporate constraints on the set of quantum states. Since the set of separable states is generally difficult to characterise (see e.g.~\cite{DPS2002}), we instead adopt an approach in which we use the ideal entanglement witness condition, $\mathcal{W}\leq \mathcal{W}_\text{sep}(0)$, which we may realistically assume to possess, in place of the set of separable states. Then, since the probabilities associated to performing the target measurements on the state explicitly appear in our moment matrix, we can introduce it as an additional linear constraint in our SDP. Hence, the optimisation is effectively a relaxation of the subset of entangled states for which the original entanglement witness holds. In fact, since the set of separable states is characterised by infinitely many linear entanglement witnesses, one can in this way continue to introduce linear standard witnesses to constrain the effective state space in the SDP and thus further improve the accuracy of the bound $\mathcal{W}_\text{sep}^\uparrow(\varepsilon)$. In Supplementary Material we exemplify the use of this method, in its basic version, using only a single witness constraint $\mathcal{W}\leq \mathcal{W}_\text{sep}(0)$ on the state space, and show that it returns non-trivial, albeit not tight, bounds for two simple entanglement witnesses for relevant values of $\varepsilon$.

%
%

\textit{Discussion.---} We have introduced and investigated entanglement detection when the measurements only nearly correspond to those intended to be performed in the laboratory. We have shown the relevance of the concept, presented explicit entanglement witnesses that take measurement inaccuracy into account, and finally shown how SDP methods can be applied to these types of problems. These results are a step towards a theoretical framework for detecting  entanglement based on devices that are quantitatively benchmarked in an operationally meaningful and experimentally accessible manner.

Our work leaves several natural open problems. If given an arbitrary standard entanglement witness, how can we compute corrections due to the introduction of measurement inaccuracies? Our SDP method is a first step towards addressing this problem but better methods are necessary both in terms of computational cost and in terms of the accuracy of the separable bound. Moreover, for a given $d$, what is the smallest number of auxiliary global measurement settings needed to eliminate the diverging derivative for optimal standard entanglement witnesses  under small measurement inaccuracy? In addition, can one extend our entanglement witnesses to witnesses of genuine higher-dimensional entanglement, e.g.~by detecting the Schmidt number?
Also, in this first work, we have focused on bipartite entanglement. It would be interesting to identify useful entanglement witnesses for multipartite states at bounded measurement inaccuracy. Finally, the framework proposed here for entanglement detection draws inspiration from ideas proposed in semi-device-independent quantum communications. Given that several frameworks for semi-device-independence recently have been proposed   \cite{VanHimbeeck2017, info1, info2, Wang2019, Tavakoli2021}, there may be other similarly inspired avenues for entanglement detection based on quantitative benchmarks.

\begin{acknowledgements}
	The authors thank Mateus Ara\'ujo for discussions. This project was  supported by the Wenner-Gren Foundations, the Austrian Science Fund (FWF) through the projects Y879-N27 (START) and P 31339-N27 (Stand-Alone),  JSPS Overseas Research Fellowships, and JST PRESTO Grant Number JPMJPR201A.
\end{acknowledgements}

\bibliography{biblio_epsilonEW}

\appendix

\section{Simplest entanglement witness}\label{AppSimple}
Consider the entanglement witness $\mathcal{W}=\expect{\sigma_X\otimes\sigma_X}+\expect{\sigma_Z\otimes\sigma_Z}$ on a pair of qubits. We allow the lab observables to have an $\varepsilon$-deviation with respect to the target measurements $\{\sigma_X,\sigma_Z\}$ on both sites. This corresponds to the constraints
\begin{align}
& \Tr\left(A_1\sigma_X\right)\geq 2-4\varepsilon, && \Tr\left(A_2\sigma_Z\right)\geq 2-4\varepsilon,\\
& \Tr\left(B_1\sigma_X\right)\geq 2-4\varepsilon, && \Tr\left(B_2\sigma_Z\right)\geq 2-4\varepsilon,
\end{align}
where we have chosen that all measurements are subject to the same magnitude of inaccuracy.

Due to the symmetry of $\mathcal{W}$ under a party swap, we can choose $A_1=B_1$ and $A_2=B_2$. Since the measurements are characterised by a pair of Bloch vectors, we can without loss of generality choose them in the $XZ$-plane of the Bloch sphere. We therefore write $A_k=B_k=\cos\theta_k \sigma_X+\sin\theta_k \sigma_Z$. In the relevant case of equality, the fidelity conditions then become 
\begin{align}
&\theta_1=-\arccos\left(1-2\varepsilon\right),\\
&\theta_2=\arcsin\left(1-2\varepsilon\right).
\end{align}
Due to the party symmetry, we can choose a product state on the form  $\ket{\phi}\otimes \ket{\phi}$ where $\ket{\phi}=\cos z\ket{0}+\sin z\ket{1}$. Then we obtain
\begin{equation}
\mathcal{W}=1+4\left(1-2\varepsilon\right)\sqrt{\varepsilon(1-\varepsilon)}\sin(4z),
\end{equation}
which is optimal at $z=\frac{\pi}{8}$ when $\varepsilon\leq \frac{1}{2}$. Hence
\begin{equation}
\mathcal{W}_\text{sep}=1+4\left(1-2\varepsilon\right)\sqrt{\varepsilon(1-\varepsilon)}.
\end{equation}
Notice that this is only valid for $\varepsilon\leq \frac{1}{2}-\frac{1}{2\sqrt{2}}$. For larger $\varepsilon$ we have $\mathcal{W}_\text{sep}=2$.

Moreover, we note that the immediate generalisation of this witness, namely $\mathcal{W}=\expect{\sigma_X\otimes\sigma_X}+\expect{\sigma_Y\otimes\sigma_Y}+\expect{\sigma_Z\otimes\sigma_Z}$, in the presence of measurement inaccuracies, can by similar means be shown to admit the separable bound
\begin{equation}
\mathcal{W}_\text{sep}=2+4\sqrt{2}\left(1-2\varepsilon\right)\sqrt{\varepsilon(1-\varepsilon)}-(1-2\varepsilon)^2,
\end{equation}
when $\varepsilon\leq \frac{3-\sqrt{3}}{6}$ and $\mathcal{W}_\text{sep}=3$ otherwise.

\section{Entanglement detection based on the CHSH quantity}\label{AppCHSH}
Consider a pair of qubits, each of which is subject to two measurements. The target observables on both sites are  $\sigma_X$ and $\sigma_Z$. The lab observables all have the same inaccuracy bound $\varepsilon$. Thus we have 
\begin{align}
& \Tr\left(A_1\sigma_X\right)\geq 2-4\varepsilon, && \Tr\left(A_2\sigma_Z\right)\geq 2-4\varepsilon,\\\label{cons}
& \Tr\left(B_1\sigma_X\right)\geq 2-4\varepsilon, && \Tr\left(B_2\sigma_Z\right)\geq 2-4\varepsilon.
\end{align}
In case of perfect measurements, the CHSH quantity acts as a conventional entanglement witness,
\begin{equation}
\mathcal{W}=\expect{A_1\otimes B_1}+\expect{A_1\otimes B_2}+\expect{A_2\otimes B_1}-\expect{A_2\otimes B_2}\leq \sqrt{2},
\end{equation}
which is respected by all separable states. Evidently, since $\mathcal{W}\leq 2$ for local hidden variable models, which in particular account for the statistics of any measurements performed on a separable state, it follows that entanglement can be detected for arbitrary $\varepsilon$. 

We show the potential influence of small measurement inaccuracies through an explicit quantum model. Choose $A_1=B_1$ and associate it to a Bloch vector $\vec{n}_1=\left(\cos \alpha,0,\sin\alpha\right)$ in the XZ-plane. Similarly choose $A_2=B_2$ and associate it to the Bloch vector $\vec{n}_2=\left(\cos \beta,0,\sin\beta\right)$. Our strategy is to align the two Bloch vectors as much as possible under the constraints \eqref{cons}. This implies the choice of 
\begin{align}
& \alpha=\arccos\left(1-2\varepsilon\right), && \beta=\arcsin\left(1-2\varepsilon\right).
\end{align}
Then, we choose the product state $\ket{\psi}=\ket{\phi}\otimes\ket{\phi}$ with $\ket{\phi}=\cos z\ket{0}+\sin z\ket{1}$, where
\begin{equation}
z=-\frac{\pi}{4}+\frac{1}{4}\arctan\left(\frac{1}{8\varepsilon-8\varepsilon^2-1}\right).
\end{equation}
The angle has been choosen so as to place the Bloch vector of $\ket{\phi}$ right in the middle of $\vec{n}_1$ and $\vec{n}_2$. This leads to the following value of the CHSH quantity,
\begin{align}
\mathcal{W}=4\left(1-2\varepsilon\right)\sqrt{\varepsilon(1-\varepsilon)}+\sqrt{2-16\varepsilon\left(1-\varepsilon\right)\left(1-2\varepsilon\right)^2},
\end{align}
when $\varepsilon\leq \frac{1}{2}-\frac{1}{2\sqrt{2}}$ and $\mathcal{W}=2$ otherwise. The derivative diverges as $\varepsilon\rightarrow 0^+$, indicating the first-order impact of small measurement inaccuracies. For small $\varepsilon$, the value scales as $\mathcal{W}\sim \sqrt{2}+4\sqrt{\varepsilon}-4\sqrt{2}\varepsilon$. For example, if we choose $\varepsilon=0.5\%$, the separable model achieves $\mathcal{W}=1.67$ which is a perturbation comparable to that obtained in the main text for the simplest two-qubit entanglement witness.

\section{Lower bounds: alternating convex search}\label{AppSeesaw}

Consider that we are given an arbitrary linear functional $\mathcal{W}$, arbitrary target measurements $\{\tilde{A}_{a|x}\}$ and $\{\tilde{B}_{b|y}\}$ and arbitrary measurement inaccuracies $\{\varepsilon^\text{A}_x,\varepsilon^\text{B}_y\}$. Consider a linear functional
\begin{equation}
\mathcal{W}=\sum_{a,b,x,y}c_{abxy} \Tr\left[A_{a|x}\otimes B_{b|y}\rho\right],
\end{equation}
with some real coefficients $c_{abxy}$. We describe a numerical method, based on alternating convex search, to systematically establish lower bounds on both $\mathcal{W}_\text{sep}$ and $\mathcal{W}_\text{ent}$. To this end we consider latter case first.

In order to place a lower bound on $\mathcal{W}_\text{ent}$, we decompose the optimisation problem into three parts: one over the measurements on system A, one over the measurements on system B and one over the global shared state. To this end, we first choose a random set of measurements $\{B_{b|y}\}$ and a random pure state $\rho$. Then, we optimise $\mathcal{W}$ over the measurements $\{A_{a|x}\}$ under the constraint that $\mathcal{F}^\text{A}_x\geq 1-\varepsilon^\text{A}_x$. This optimisation is a semidefinite program and can therefore be efficiently solved. Using the returned measurements $\{A_{a|x}\}$, we optimise $\mathcal{W}$ over the measurements $\{B_{b|y}\}$ under the constraint that $\mathcal{F}^\text{B}_y\geq 1-\varepsilon^\text{B}_y$. This is again a semidefinite program. Finally, using the returned measurements $\{B_{b|y}\}$, we evaluate the Bell operator
\begin{equation}
\mathcal{B}=\sum_{a,b,x,y}c_{abxy}A_{a|x}\otimes B_{b|y}
\end{equation}
and compute its largest eigenvalue. The associated eigenvector is the optimal state, which corresponds to our choice of $\rho$. This routine of two semidefinite programs and one eigenvalue computation can then be iterated in order to find increasingly accurate lower bounds on $\mathcal{W}_\text{ent}$. The procedure depends on the initial starting point and ought therefore to be repeated several times independently.

To place a lower bound on  $\mathcal{W}_\text{sep}$, we can proceed analogously to the above when treating the separate optimisations over the measurements $\{A_{a|x}\}$ and $\{B_{b|y}\}$. However, the optimisation over the state is now less straightforward since we require that $\rho=\ketbra{\phi}{\phi}\otimes \ketbra{\psi}{\psi}$. The optimisation over the state can be cast as another alternating convex search, treated as a sub-routine to the main alteranting convex search. In other words, we sample a random $\ket{\phi}$ and evaluate the semidefinite program optimising $\mathcal{W}$ over $\ket{\psi}$. Then, using the returned $\ket{\psi}$, we run a semidefinite program optimising $\mathcal{W}$ over $\ket{\phi}$. This procedure is iterated until desired convergence is obtained.

\section{Bounds on witness}\label{AppWitness}
Let $\{\lambda_i\}_{i=1}^{d^2-1}$ be an orthonormal basis the space of operators acting on $d$-dimensional Hilbert space, with $\Tr\left(\lambda_i\lambda_j^\dagger\right)=d\delta_{ij}$.  Then, every qudit state can be written as
\begin{equation}\label{bloch}
\rho=\frac{1}{d}\left(\openone+\vec{\mu}\cdot\vec{\lambda}\right),
\end{equation}
where $\vec{\mu}$ is some complex-valued Bloch vector with entries $\mu_i=\expect{\lambda_i^\dagger}=\Tr\left(\rho \lambda_i^\dagger\right)$. By checking the purity $\Tr\left(\rho^2\right)$, one finds that $\norm{\vec{\mu}}^2=\sum_{i=1}^{d^2-1}\expect{\lambda_i^\dagger}^2\leq d-1$. In general, not every such Bloch vector corresponds to a valid density matrix. 

Consider the witness
\begin{equation}
\mathcal{W}^{(d)}=\sum_{i=1}^n \expect{\lambda_i\otimes \bar{\lambda}_i}.
\end{equation}

For separable states, we can evaluate $\mathcal{W}^{(d)}_\text{sep}$ by restricting to product states. Then we have
\begin{align}\nonumber\label{C3}
\mathcal{W}^{(d)}&=\sum_{i=1}^n \expect{\lambda_i}_A \expect{\bar{\lambda}_i}_B\leq \sqrt{\sum_{i=1}^n  \expect{\lambda_i}^2_A}\sqrt{\sum_{i=1}^n \expect{\bar{\lambda}_i}_B^2}\\
&\leq d-1 = \mathcal{W}^{(d)}_\text{sep}.
\end{align}
Notice that this is independent of $n$.

For entangled states, we have 
\begin{align}
&\mathcal{W}^{(d)}\leq \sum_{i=1}^n \expect{\lambda_i\otimes \bar{\lambda}_i}=\nu_\text{max}\left[\sum_{i=1}^n \lambda_i\otimes \bar{\lambda}_i \right]\\
&\leq n \max_i \nu_\text{max}\left[\lambda_i\otimes \bar{\lambda}_i\right]=n \max_i \nu_\text{max}\left[\lambda_i\right]^2\leq n(d-1),
\end{align}
where we used that $\nu_\text{max}\left[\lambda_i\right]\leq \sqrt{d-1}$. However this, essentially trivial, bound is only tight for $d=2$, in which case it is algebraically maximal. To obtain a bound for $d>2$, we note that the entangled state lives in dimension $d^2$. Hence, its Bloch vector length is at most $\sqrt{d^2-1}$. In other words,
\begin{equation}
\sum_{i=1}^{n} \expect{\lambda_i\otimes \bar{\lambda}_i}^2\leq d^2-1.
\end{equation}
Taking the case of equality, we obtain a bound on the largest value of the witness when all entries in the sum are equal. Thus we require 
\begin{equation}
\expect{\lambda_i\otimes \bar{\lambda}_i}=\sqrt{\frac{d^2-1}{n}},
\end{equation}
which gives
\begin{equation}
\mathcal{W}^{(d)}_\text{ent}\leq  \sqrt{n}\sqrt{d^2-1}.
\end{equation} 
This bound is not necessarily tight.

Consider now the case when we have separable states and inaccurate measurements. Expand $\mathcal{W}^{(d)}$ as follows,
\begin{align}\nonumber
\mathcal{W}^{(d)}=& \sum_{i=1}^n\expect{A_i\otimes B_i}=q^2\sum_{i=1}^n\expect{\lambda_i}_A\expect{\bar{\lambda}_i}_B\\\nonumber
&+q\sqrt{1-q^2}\sum_{i=1}^n\left(\expect{\lambda_i}_A\expect{\bar{\lambda}_{i}^\perp}_B+\expect{\lambda_{i}^\perp}_A\expect{\bar{\lambda}_i}_B\right)\\
&+\left(1-q^2\right)\sum_{i=1}^n\expect{\lambda_{i}^\perp}_A\expect{\bar{\lambda}_{i}^\perp}_B.
\end{align}
We examine these sums one by one. From \eqref{C3}, we see that the first sum is at most $d-1$. Next, we use the Cauchy-Schwarz inequality to write the second sum as
\begin{align}\nonumber
\sum_{i=1}^n& \expect{\lambda_i}_A\expect{\bar{\lambda}_{i}^\perp}_B\leq\sqrt{\sum_{i=1}^n\expect{\lambda_i}_A^2}\sqrt{\sum_{i=1}^n\expect{\bar{\lambda}_{i}^\perp}_B^2}\\
&\leq \sqrt{d-1}\sqrt{\sum_{i=1}^n\expect{\bar{\lambda}_{i}^\perp}_B^2}\leq \left(d-1\right)\sqrt{n-1}.
\end{align}
In the last step, we have used the following lemma. Let $\vec{u}\in\mathbb{R}^n$ and $\vec{v}^i\in\mathbb{R}^n$ be unit vectors such that the $i$'th component of $\vec{v}^i$ is zero, i.e.~$\vec{v}_i^i=0$. Then we have that
\begin{equation}
\sum_{i=1}^n \left(\vec{u}\cdot \vec{v}^i\right)^2\leq \sum_{i=1}^n 1-\vec{u}_i^2=n-1.
\end{equation}
Again using the Cauchy-Schwarz inequality and this lemma also leads to
\begin{align}
&\sum_{i=1}^n\expect{\lambda_{i}^\perp}_A\expect{\bar{\lambda}_{i}}_B\leq \left(d-1\right)\sqrt{n-1},\\
&\sum_{i=1}^n\expect{\lambda_{i}^\perp}_A\expect{\bar{\lambda}_{i}^\perp}_B \leq \left(d-1\right)\left(n-1\right).
\end{align}
Putting it together, we arrive at the bound
\begin{equation}\label{res}
\mathcal{W}_\text{sep}\leq \left(d-1\right)\left(n-1-q^2\left(n-2\right)+2q\sqrt{n-1}\sqrt{1-q^2}\right).
\end{equation}

\section{Semidefinite relaxations}\label{AppSDP}
Consider the task of optimising an arbitrary linear functional over the set of projective quantum strategies with a given inaccuracy to a set of target measurements:
\begin{align}\nonumber\label{optim}
&	\qquad \qquad \mathcal{W}_\text{ent}=\max_{\{A_{a|x}\},\{B_{b|y}\},\rho} \mathcal{W}[p] \\\nonumber
& \text{subject to }\quad \Tr\left(\rho\right)=1, \qquad \rho\geq 0, \qquad \rho\in \mathcal{L}(\mathbb{C}^d)\\ \nonumber
& A_{a|x}A_{a'|x}=A_{a|x}\delta_{a,a'}, \qquad B_{b|y}B_{b'|y}=B_{b|y}\delta_{b,b'},\\\nonumber
& \sum_a A_{a|x}=\openone_d, \qquad \sum_b B_{b|y}=\openone_d\\\nonumber
& \mathcal{F}^\text{A}_x\geq 1-\varepsilon^\text{A}_x, \qquad  \mathcal{F}^\text{B}_y\geq 1-\varepsilon^\text{B}_y\\
& p(a,b|x,y)=\Tr\left[A_{a|x}\otimes B_{b|y} \rho\right],
\end{align}
where $\mathcal{L}(\mathbb{C}^d)$ is the set of linear operators of dimension $d$.  This is generally a difficult optimisation problem. However, it can be relaxed into a hierarchy of increasingly precise criteria, each of which can be evaluated as a semidefinite program. 

To this end, define the operator list
\begin{equation}
	S=\{\openone_{d^2}, \rho, \{A_{a|x}\}_{a,x},\{B_{b|y}\}_{b,y},\{\tilde{A}_{a|x}\}_{a,x},\{\tilde{B}_{b|y}\}_{b,y}\}.
\end{equation}
Here, the measurement operators are to be understood as spanning the full Hilbert space, e.g.~$A_{a|x}\rightarrow A_{a|x}\otimes \openone_d$. We let $M_k$ denote the set of all monomials, taken from the list $S$, of degree at most $k$. We let $n(k)$ denote the size of the set $M_k$. Then, we define the $n(k)\times n(k)$ tracial moment matrix as
\begin{equation}
	\Gamma(u;v)=\Tr\left(uv^\dagger\right),
\end{equation}
for $u,v\in M_k$. A quantum model implies the positivity of $\Gamma$. Moreover, by including enough monomials, we can formulate the objective  as a linear function in the moment matrix,
\begin{equation}\label{obj}
	\mathcal{W}(\Gamma)=\sum_{a,b,x,y} c_{abxy} \Gamma(\rho A_{a|x}; B_{b|y}).
\end{equation}
Similarly, the inaccuracy constraints can be formulated as the linear constraints
\begin{align}\nonumber\label{sdpcons}
&	\frac{1}{d^2}\sum_{a=1}^o \Gamma(A_{a|x}; \tilde{A}_{a|x})\geq 1-\varepsilon^\text{A}_x, \\
& \frac{1}{d^2}\sum_{b=1}^o \Gamma(B_{b|y}; \tilde{B}_{b|y})\geq 1-\varepsilon^\text{B}_y. 
\end{align}

In order to capture the constraints of $d$-dimensional Hilbert space and to fix the target measurements in the optimisation, we proceed as follows \cite{Navascues2015, Tavakoli2021}. We randomly sample $\rho$, $\{A_{a|x}\}_{a,x}$ and $\{B_{b|y}\}_{b,y}$ from a $d$-dimensional Hilbert space and construct the list $S$. Note that the target measurements are fixed at all times. Then, we evaluate the moment matrix and label it $\Gamma^{(1)}$. This process is repeated, leading to a list of sampled moment matrices $\{\Gamma^{(1)},\ldots,\Gamma^{(m)}\}$. The sampling is terminated when the next moment matrix is found to be linearly dependent on all the previously sampled moment matrices. Thus, the sampled list constitutes a (non-orthonormal) basis of the space of moment matrices. We then define the total moment matrix as the affine combination
\begin{align}\label{sdpcons2}
& \Gamma=\sum_{i=1}^m s_i \Gamma^{(i)}, && \sum_{i=1}^m s_i=1,
\end{align}
where $\{s_i\}$ serve as optimisation variables. 

We can now formulate our relaxation of the optimisation problem \eqref{optim} as $\mathcal{W}_\text{ent}(\varepsilon)\leq \mathcal{W}_\text{ent}^\uparrow(\varepsilon) $ where
\begin{align}
& \mathcal{W}_\text{ent}^\uparrow \equiv \max_{\{s_i\}} \mathcal{W}(\Gamma) \quad \text{ subject to } \quad \Gamma\geq 0
\end{align}
under the constraints \eqref{sdpcons} and \eqref{sdpcons2}. This can be evaluated as a semidefinite program. The relaxation becomes tighter as the list of monomials $M_k$ is extended.

\begin{figure}
	\centering
	\includegraphics[width=\columnwidth]{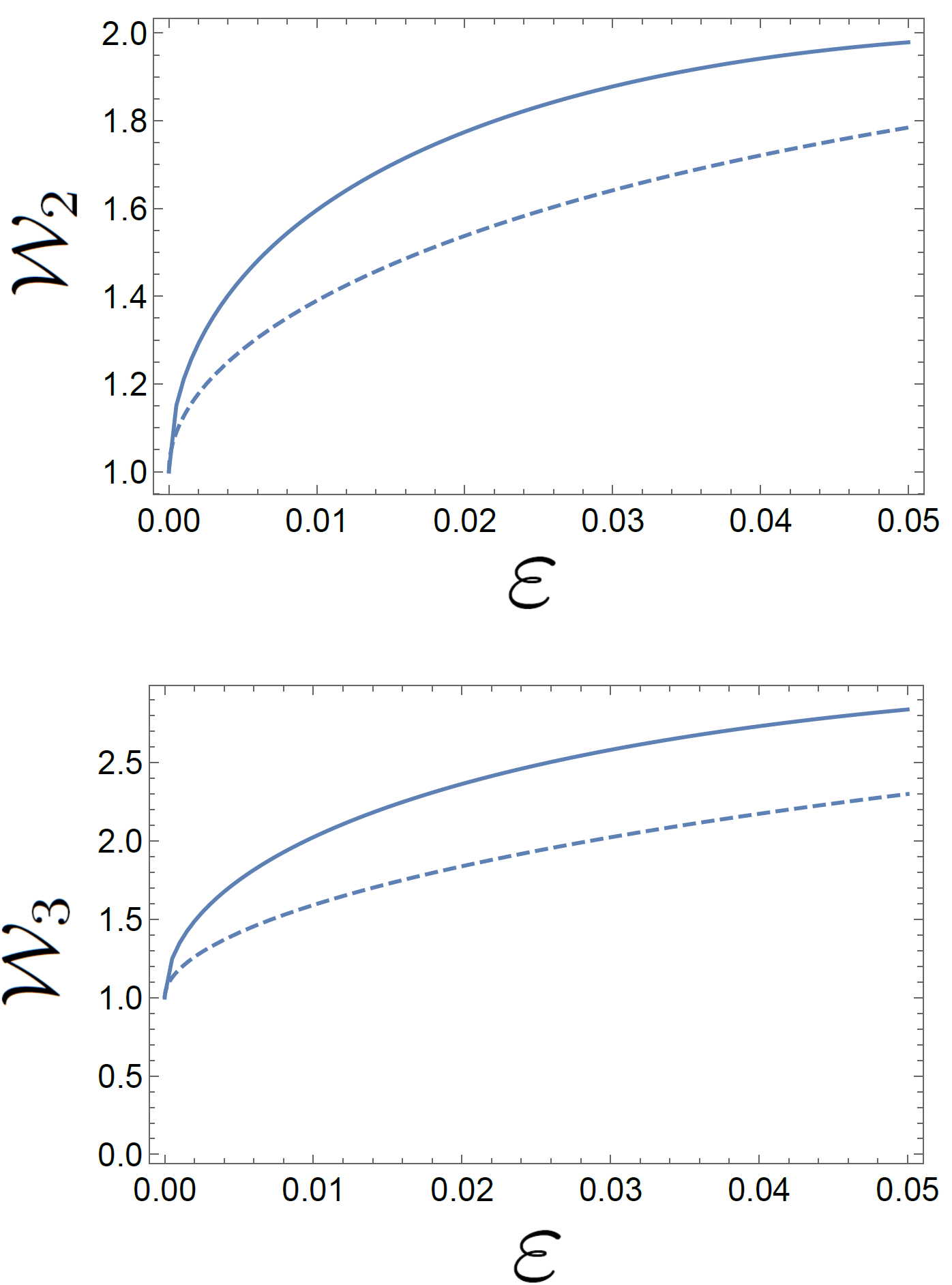}
	\caption{\textbf{Solid lines.} Bounds $ \mathcal{W}_\text{sep}^\uparrow(\varepsilon)$ obtained for the witness $\mathcal{W}_2$ and $\mathcal{W}_3$ via SDP relaxations where separability is relaxed to the set of entangled states obeying $\mathcal{W}_2(0)\leq 1$ and $\mathcal{W}_3(0)\leq 1$ respectively. These bounds can likely be made tighter by adding more ideal entanglement witness constraints to the SDP, in order to improve the relaxation of separability. \textbf{Dashed lines.} Optimal separable bound calculated analytically in Appendix~\ref{AppSimple}.}\label{Fig_SDP}
\end{figure}

In order to instead obtain bounds of the form  $\mathcal{W}_\text{sep}(\varepsilon)\leq \mathcal{W}_\text{sep}^\uparrow(\varepsilon) $, we can add the constraint 
\begin{equation}\label{ew}
	\sum_{a,b,x,y} c_{abxy}\Gamma(\rho \tilde{A}_{a|x}; \tilde{B}_{b|y})\leq \mathcal{W}_\text{sep}(0),
\end{equation}
which corresponds to a standard entanglement witness. Note that we can introduce even more ``target'' measurements in the operator list $S$, thus extending the size $n(k)$ of the moment matrix, and then use them to build additional linear constraint like \eqref{ew} representing standard entanglement witnesses. The introduction of these shrinks the effective state space, thus improving the accuracy of the bound $ \mathcal{W}_\text{sep}^\uparrow(\varepsilon)$, at the price of a larger SDP.

We exemplify a  simple version of this method for the case of the two witnesses considered in Appendix~\ref{AppSimple}, namely $\mathcal{W}_2=\expect{\sigma_X\otimes\sigma_X}+\expect{\sigma_Z\otimes\sigma_Z}\leq 1$ and $\mathcal{W}_3=\expect{\sigma_X\otimes\sigma_X}+\expect{\sigma_Y\otimes\sigma_Y}+\expect{\sigma_Z\otimes\sigma_Z}\leq 1$, at inaccuracy $\varepsilon$. These are evaluated with monomial lists of length $46$ and $89$ respectively. The results are illustrated in Figure~\ref{Fig_SDP}. As expected, the returned bounds are not tight, due to the basic relaxation of the separable set to all entangled states obeying $\mathcal{W}\leq 1$. Nevertheless, the bounds are non-trivial for relevant values of $\varepsilon$.

\end{document}